\documentclass[prl,aps,twocolumn,showpacs,showkeys,superscriptaddress]{revtex4} %no pics displayed

\usepackage{epsfig,amssymb,amsfonts,verbatim}

\def\bfl{\begin{flushleft}}
\def\efl{\end{flushleft}}
\def\bfr{\begin{flushright}}
\def\efr{\end{flushright}}
\def\bc{\begin{center}}
\def\ec{\end{center}}
\def\be{\begin{equation}}
\def\ee{\end{equation}}
\def\ba{\begin{eqnarray}}
\def\ea{\end{eqnarray}}
\def\baa#1{\begin{array}{#1}}
\def\eaa{\end{array}}
\def\bw{\begin{widetext}}
\def\ew{\end{widetext}}

\def\text#1{\mbox{#1}}

\begin{document}

%\preprint{APS/123-QED}

\title{Hall-Lorenz ratio of YBa$_2$Cu$_3$O$_7$ using Ionization
energy based Fermi-Dirac statistics and charge-spin separation}

\author{
Andrew Das Arulsamy}
%\cite{z}

\affiliation{Condensed Matter Group, Division of Exotic Matter,
No. 22, Jalan Melur 14, Taman Melur, 68000 Ampang, Selangor DE,
Malaysia}

%\date{1$^{st}$ January 2003}
%\date{~Received: 12 Jan 2001 [LANL] ~}
%\date{~Received: 26 May 2000 [PRL], 1 June 2000 [LANL] ~}
\date{\today}


%\scriptsize%\footnotesize

\begin{abstract}
The temperature dependent properties of heat capacity, heat
conductivity and Hall-Lorenz ratio have been solved numerically
after taking the previously proposed ionization energy based
Fermi-Dirac statistics and the coexistence of Fermi and
charge-spin separated liquid into account. The
thermo-magneto-electronic properties are entirely for spin and
charge carriers, hence the phonon contribution has been neglected.
A linear dependence between the Hall-Lorenz ratio and temperature
($T$) is also obtained in accordance with the experimental results
for overdoped YBa$_2$Cu$_3$O$_{7-\delta}$, if these conditions,
$E_I$ $<$ $T_c$ with respect to resistivity and there are no
spinon pairings ($T^*$ = 0) are satisfied. Heat conductivity based
on both pure and electron-contaminated charge-spin separated
liquid in $ab$-planes above $T_c$ are found to increase with
decreasing $T$ as a consequence of inverse proportionality with
$T$. The $T$-dependence of heat capacity are also highlighted,
which qualitatively complies with the experimental findings.

\end{abstract}


\maketitle

%\narrowtext

%\small

\section{1. Introduction}

The mechanism of superconductivity in YBa$_2$Cu$_3$O$_{7-\delta}$
has got to be of both electronic and magnetic in nature. However,
its transport properties in the normal state (above $T_c$) of
high-$T_c$ superconductors (HTSC) are still clouded with
obscurities due to anisotropy and strong correlations. On the one
hand, there is a lack of confluence on the interpretations of
ARPES (Angle-Resolved Photoemission Spectroscopy) and scanning
tunnelling spectroscopic (STS) measurements convincingly with
respect to superconductor, spin and charge gaps and its
evolutions, as well as the holon condensation and precursor
superconductivity (Cooper pairs above
$T_c$)~\cite{krasnov2,kleefisch,damascelli,kaminski}. On the other
hand, the charge and spin carriers in the normal state seem to
obey both Fermi and non-Fermi
liquid~\cite{maekawa,timusk,takeda,batlogg} somewhat mimicking the
duality property of light (particle-wave). Instead of assuming
that there are no such connections of which the normal state
properties are somehow influenced by the types of experiments, one
can entirely and safely neglect this conception by heavily relying
on the basic transport experiments such as the resistivity, Hall
resistance, Hall-Lorenz ratio, heat capacity and conductivity.
Anyhow, the normal state properties with characteristics or spin
gap temperature ($T^*$) and $T_{crossover}$ must consist of at
least two types of strongly interacting liquid. Theoretical
studies based on other experimental techniques, such as Optical,
STS, ARPES and NMR (Nuclear Magnetic Resonance) are also essential
and need to be carried out in order to extract the missing
puzzles, which could act as a bridge between the theoretical
interpretations based on transport experiments and other
experimental techniques.

In addition, the former focus-point based on basic transport
experiments also suggests that not all magneto-electronic
properties and its magnitudes determined from the STS, optical,
NMR and ARPES measurements are directly applicable for the basic
transport mechanisms and vice versa. This scenario arises as a
result of the principle of least action that always plays a
crucial and dominant role in any transport
measurements~\cite{arulsamyssc}, which is actually reinforced by
the level of anisotropy. As an example, the magnitude of energy
gaps for ferromagnetic manganites above Curie temperature
(paramagnetic phase), doped semiconductors with limited solubility
of the dopants and even in semiconducting normal states of
superconductors determined from the resistivity measurements are
always smaller than the ones measured and/or calculated from other
experimental techniques, including STS. Consequently, the intense
focus here will be given on the basic transport experiments in
order to explain the thermo-magneto-electronic properties of
YBa$_2$Cu$_3$O$_{7-\delta}$ theoretically. It is well known that
the popular theoretical proposals like, stripes, pure charge-spin
separation (CSS), polaron-bipolaron model, quantum critical points
and marginal Fermi
liquid~\cite{batlogg,varma,timusk,andersoni,andersonii,leep,sachdev,tranquada,alexi,alexii,maekawa}
seriously lack the ability in one way or another to predict all
the transport properties consistently and convincingly in the
presence of electric field, magnetic field and heat gradient or
any of its combinations in the temperature range between $T_c$ and
300 K. Instead of considering the pure CSS liquid, CSS in the
midst of electrons is used here to comply, first with all the
basic transport measurements consistently before embarking on
other experimental results. The reason for choosing CSS related
mechanism is that it works extremely well qualitatively for the
resistivity and Hall coefficient-angle measurements at $T$ ranging
from 300 K $\to$ $T_c$, taking care of both $T^*$ and
$T_{crossover}$ effects.

In the early developments for the theory of HTSC, Lee and
Nagaosa~\cite{lee} have predicted the violation of Wiedemann-Franz
law ($\kappa\rho/T$; $\kappa$ and $\rho$ denote the heat
conductivity and electrical resistivity respectively)
theoretically based on pure CSS. This breakdown was thought to be
solely as a result of incompatibility between the heat and charge
conductivity rules, which turns out to be insufficient and
additional constraints are required~\cite{arulsamy5}. The charge
conductivity obeys the parallel-circuit's resistance rule,
presently known as the Ioffe-Larkin rule~\cite{ioffe} whereas the
heat conductivity follows the series-circuit's resistance
rule~\cite{lee}. Unfortunately, CSS has been found to be
energetically unfavorable in 2-dimensional (2D) systems, unlike in
1D systems as pointed out by Sarker~\cite{sarker}. Sarker showed
that pure CSS liquid gives rise to excessive kinetic energy
relative to lowest possible kinetic energy, which will eventually
prohibit stable recombination processes for the occurrences of
superconductivity in 2D and Fermi liquid in 3D. Nevertheless, CSS
that leads to spinons and holons as spin and charge carriers
respectively were allowed to coexist with electrons in $ab$-planes
recently so as to fulfill the dimensionality
crossover~\cite{arulsamy1,arulsamy2,arulsamy3} as well as to avoid
this excessive kinetic energy scenario. In Ref.~\cite{arulsamy3},
note that there are three typographical errors in this paper, two
of them can be found in Eq.~(9) where $(m^*_em^*_h)^{-3/4}$ and
$(2\pi \hbar^2/k_B)^{-3/2}$ should be replaced with
$(m^*_em^*_h)^{-1/4}$ and $(2\pi \hbar^2/k_B)^{3/2}$ respectively.
Therefore, the fitting parameter, $A_2$ is actually equals to
$(A_2/2e^2)(m^*_em^*_h)^{-1/4}(2\pi \hbar^2/k_B)^{3/2}$. The
electrons here obey the ionization energy based Fermi-Dirac
statistics~\cite{arulsamyi,arulsamyii} (iFDS). A hybrid model
based on this coexistence of Fermi and charge-spin separated
(FCSS) liquid was proposed by employing the Ichinose-Matsui-Onoda
(IMO) equation~\cite{ichinose4} that has been successful to
explain both $c$-axis and $ab$-plane's resistivities ($\rho_c(T)$,
$\rho_{ab}(T)$) as well as its dimensionality crossover including
the Hall resistances~\cite{arulsamy5} ($R_H^{(c)}$, $R_H^{(ab)}$).
In the presence of magnetic field ({\bf H}), FCCS liquid also
obeys the Anderson's hypothesis~\cite{anderson6},
1/$\tau_{transport}$ $\propto$ $T$ and 1/$\tau_{Hall}$ $\propto$
$T/\tau_{transport}$ $\propto$ $T^2$. Likewise, it is stressed
here that the thermo-magneto-electronic properties of
YBa$_2$Cu$_3$O$_{7-\delta}$ can also be explained using the FCSS
liquid without any additional assumptions or constraints, apart
from the ones given in Ref.~\cite{arulsamy5}. However, note that
the thermal related physical parameters derived here neglect the
phonon contribution.

\section{2. Theoretical Details}
\subsection{2.1. Heat capacity and related distribution functions}

Holon (spinless charged-boson), spinon (chargeless fermion) and
electron can carry heat (entropy). The respective particles can be
excited to a higher energy levels satisfying the Bose-Einstein
statistics ($f_{BES}(E)$), Fermi-Dirac statistics ($f_{FDS}(E)$)
and ionization energy based Fermi-Dirac statistics
($f_{iFDS}(E)$). Therefore, the heat capacity in $ab$-planes can
be explicitly written as

\begin{eqnarray}
&\mathcal{C}_{ab}& = DOS\Bigg[\int^\infty_0 (E-E_F)\frac{\partial
f_{BES}(E)}{\partial T}dE \nonumber
\\&& ~~+ \int^\infty_{\Delta_{SG}} (E-E_F)\frac{\partial f_{FDS}(E)}{\partial
T}dE \nonumber \\&& ~~+ \int^\infty_0 (E-E_F-E_I)\frac{\partial
f_{iFDS}(E)}{\partial T}dE \Bigg] \nonumber \\&& =
\frac{1}{\pi\hbar^2}\bigg[m_s\Phi_{FDS}(T) \nonumber
\\&& ~~+ m_h\Phi_{BES}(T) +
m_e^*\Phi_{iFDS}(T)\bigg]. \label{eq:1}
\end{eqnarray}

The $E$-independent 2D density of states is, $DOS$ =
$m^*/\pi\hbar^2$. $\hbar$ = $h/2\pi$, $h$ denotes Planck constant,
$k_B$ is the Boltzmann constant while $m^*$ represents the
effective mass. $\Phi_{BES}(T)$, $\Phi_{FDS}(T)$ and
$\Phi_{iFDS}(T)$ denote the integrals with respect to BES, FDS and
iFDS respectively. In the case of pure CSS liquid, the term,
$m_e^*\Phi_{iFDS}(T)$ $\to$ 0. The respective distribution
functions for BES, FDS and iFDS are given by $f_{BES}(E) =
1/[\exp\big[(E-E_F)/k_BT\big] - 1]$, $f_{FDS}(E) =
1/[\exp\big[(E-E_F)/k_BT\big] + 1]$
and~\cite{arulsamy1,arulsamy2,arulsamyii} $f_{iFDS}(E) =
1/[\exp\big[(E-E_F+E_I)/k_BT\big] + 1]$. $E_I$ is the ionization
energy, $E_I = e^2/8\pi\epsilon\epsilon_0r_B$ and $\Delta_{SG}$ =
$T^*$, denotes the spin gap in $ab$-planes only. $\epsilon$ and
$\epsilon_0$ are the dielectric constant and permittivity of free
space respectively, whereas $e$ and $r_B$ represent the charge of
an electron and the Bohr radius respectively. Importantly, the
variation of $E_I$ with {\bf H} indicates that $r_B$ varies
accordingly with {\bf H}. Identical relationship was also given
between polaronic radius, $r_p$ and polaronic hopping energy,
$E_p$ by Banerjee {\it et al}.~\cite{banerjee}. The crucial issue
here is that $E_I$ implicitly represent the polaronic effect. Now
switching back to the heat capacity, note that the relation,
$\mathcal{C}_{ab}$ = $\mathcal{C}^s$ + $\mathcal{C}^h$ +
$\mathcal{C}^e$ directly comes from $\kappa$ = $\kappa_s$ +
$\kappa_h$ + $\kappa_e$ that follows from the total heat current
in $\kappa_{ab} = -\sum_{\alpha}j_Q^{\alpha}\nabla T$ =
$\sum_{\alpha}\mathcal{C}^{\alpha}v_F^2\tau_{\nu}/2$, $\alpha$ =
spinon (\textit{s}), holon (\textit{h}), electron (\textit{e}) and
$\nu$ = transport ($ab(s,h)$), electron ($e$).

\subsection{2.2. Heat conductivity and Hall-Lorenz ratio}

The 2D non-phononic heat conductivity in $ab$-planes,
$\kappa_{ab}(T)$ is given below after taking $E_F =
\frac{1}{2}m^*v_F^2$, $\tau_{ab(s,h)}$ (spinons and holons
interactions in $ab$-planes, neglecting spinon-pairing) and
$\tau_e$ (electron-electron interactions) into account.

\begin{eqnarray}
&&\kappa_{ab} = \kappa_s + \kappa_h + \kappa_{s + h \to e} =
\kappa_s + \kappa_h + \gamma\kappa_e \nonumber
\\&&
= \frac{E_F}{\pi\hbar^2}\bigg[\tau_{ab(h)}\Phi_{BES}(T) +
\tau_{ab(s)}\Phi_{FDS}\big(T,\lim_{\Delta_{SG}\to 0}\big)
\nonumber
\\&& ~~~+ \gamma\tau_e\Phi_{iFDS}(T)\bigg] \nonumber \\&&
= \frac{E_F}{\pi\hbar^2}\bigg[\frac{\Phi_{BES}(T)}{B_hT} +
\frac{\Phi_{FDS}(T)}{B_sT^{4/3}} +
\gamma\frac{\Phi_{iFDS}(T)}{A_eT^2}\bigg]. \label{eq:2}
\end{eqnarray}

The $ab$-plane scattering rate for the spinons and holons are
respectively given by~\cite{lee,ichinose4} $\tau_{ab(s)}$ =
$1/B_sT^{4/3}$ and $\tau_{ab(h)}$ = $1/B_hT$. $E_F$ and $v_F$
denote Fermi energy and Fermi velocity respectively whereas
$\gamma$ is the constant of
proportionality~\cite{arulsamy3,arulsamy5} that represents the
contribution of electrons in $ab$-planes. $m_s$, $m_h$ and $m^*_e$
are the holon, spinon and electron's effective mass respectively.
$B_s$, $B_h$ and $A_e$ are the spinon, holon and electron's
scattering rate dependent constants respectively (independent of
$T$) in the presence of $\nabla T$. Unlike in the previous
electrical resistivity derivations~\cite{ichinose4,arulsamy5},
spinon pairing has been neglected in this subsequent heat
conductivity calculations since $\tau_{ab}$ in the presence of
spinon-pairing and $\nabla T$ is still unclear. In other words,
the IMO's $\tau_{ab(h)}$, $1/\tau_{ab(h)}$ = $BT[1 - C(T^* -
T)^d]$ in the midst of spinon-pairing may not represent the above
$\tau_{ab(h)}$ correctly. As such, the integral, $\Phi_{FDS}(T)$
now reads as $\Phi_{FDS}(T,\lim_{\Delta_{SG}\to 0}) =
\int^\infty_0 (E-E_F)\frac{\partial f_{FDS}(E)}{\partial T} dE$.
Note that all those integrals were solved numerically. It is also
worth mentioning that even if one assumes, $\exp[(E-E_F)/k_BT]$
$\gg$ 1 and $\exp[(E-E_F+E_I)/k_BT]$ $\gg$ 1, the results remain
consistent with the numerical ones (without such assumptions) as
anticipated. Subsequently, the heat conductivity for the pure CSS
liquid can be written as

\begin{eqnarray}
&&\kappa_{ab} = \kappa_h + \kappa_s \nonumber
\\&&
= \frac{E_F}{\pi\hbar^2}\bigg[\frac{\Phi_{BES}(T)}{B_hT} +
\frac{\Phi_{FDS}(T,\lim_{\Delta_{SG}\to 0})}{B_sT^{4/3}}\bigg].
\label{eq:3}
\end{eqnarray}

On the other hand, the $c$-axis heat conductivity, $\kappa_c(T)$
is simply given by

\begin{eqnarray}
&&\kappa_c = \kappa_{e \rightleftharpoons s + h} + \kappa_e =
\beta(\kappa_h + \kappa_s) + \kappa_e \nonumber
\\&&
= \frac{E_F}{\pi\hbar^2}\bigg[\beta\tau_{ab(h)}\Phi_{BES}(T) +
\beta\tau_{ab(s)}\Phi_{FDS}\big(T,\lim_{\Delta_{SG}\to 0}\big)
\nonumber
\\&& ~~~+ \tau_e\Phi_{iFDS}(T)\bigg] \nonumber \\&&
= \frac{E_F}{\pi\hbar^2}\bigg[\beta\frac{\Phi_{BES}(T)}{B_hT} +
\beta\frac{\Phi_{FDS}(T)}{B_sT^{4/3}} +
\frac{\Phi_{iFDS}(T)}{A_eT^2}\bigg]. \label{eq:4}
\end{eqnarray}

Here, $\beta$ is the constant of
proportionality~\cite{arulsamy3,arulsamy5} that represents the
contribution of spinon-holon in $c$-axis. As a matter fact, it is
stressed that the origin of $\gamma$ and $\beta$ in
Eqs.~(\ref{eq:2}) and~(\ref{eq:4}) are as a result of the
following definitions, (which is in compliance with the mechanism
for the resistivity~\cite{arulsamy5}) the term $\kappa_{s + h \to
e}$ is defined to be the heat conductivity reduced by the process
$s$ + $h$ $\to$ $e$ occurring in the $ab$-planes. Alternatively,
it is the heat conductivity reduced by the electrons in
$ab$-planes. If $s$ + $h$ $\to$ $e$ is completely blocked in
$ab$-planes then $\kappa_{s + h \to e}$'s contribution is zilch.
Any increment in $\kappa_e$ also increases $\kappa_{s + h \to e}$
therefore $\kappa_{s + h \to e}$ = $\gamma\kappa_e$. In contrast,
the term $\kappa_{e \rightleftharpoons s + h}$ is defined to be
the heat conductivity reduced by the blockage in the process, $e$
$\rightleftharpoons$ $s$ + $h$ or the reduction in heat
conductivity caused by the blockage faced by electrons to enter
the $ab$-planes ($e$ $\to$ $s$ + $h$) and the blockage faced by
spinons and holons to leave the $ab$-planes ($s$ + $h$ $\to$ $e$).
These blockages originate from the non-spontaneity conversion of
$e$ $\rightleftharpoons$ $s$ + $h$. The reduction in heat
conductivity, $\kappa_{e \rightleftharpoons s + h}$ can also be
solely due to the blockage of $e$ $\to$ $s$ + $h$ or $s$ + $h$
$\to$ $e$. Actually, if the magnitude of blockage in $e$ $\to$ $s$
+ $h$ $>$ $s$ + $h$ $\to$ $e$ then the blockage of $e$ $\to$ $s$ +
$h$ contributes to $\kappa_c$. In short, if one of the conversion,
say $e$ $\to$ $s$ + $h$ is less spontaneous than $s$ + $h$ $\to$
$e$, then the former conversion determines the $\kappa_{e
\rightleftharpoons s + h}$. Moreover, reduction in $\kappa_{ab}$
further blocks $e$ $\rightleftharpoons$ $s$ + $h$ that leads to a
reduction in $\kappa_{e \rightleftharpoons s + h}$ hence,
$\kappa_{e \rightleftharpoons s + h}$ = $\beta\kappa_{ab}$. I.e,
the process $e \rightleftharpoons s + h$ becomes increasingly
difficult with reduction in $\kappa_{ab}$. This proportionality
can also be interpreted as the additional scattering for the
electrons to transfer heat directionally across $ab$-planes. If
$e$ $\rightleftharpoons$ $s$ + $h$ is spontaneous then the term,
$\kappa_{e \rightleftharpoons s + h}$ is null. Subsequently, the
Hall-Lorenz number is defined as $L^{(ab)}_H$ =
$\rho^{H}_{ab}(T)k_{ab}(T)/T$. As such, $L^{(ab)}_H$ can be
expressed as

\begin{eqnarray}
&L^{(ab)}_H& = \frac{1}{T}\bigg[\frac{m_h}{n_he^2\tau_{tr}}<T> +
\frac{\gamma m_e^*}{n_ee^2\tau_e} \bigg] \times \nonumber \\&&~~~
\frac{E_F}{\pi\hbar^2}\bigg[\tau_{ab(h)}\Phi_{BES}(T) +
\tau_{ab(s)}\Phi_{FDS}(T,\lim_{\Delta_{SG}\to 0}) \nonumber
\\&& ~~~+ \gamma\tau_e\Phi_{iFDS}(T)\bigg] \nonumber
\\&& = \bigg[B_{LH}\frac{m_h}{e^2n_h}T
+ \gamma
A_{LH}\frac{\pi\hbar^2}{k_Be^2}\exp\left(\frac{\Delta_{PG}}{T}\right)
\bigg] \times \nonumber \\&&
~~~~\frac{E_F}{\pi\hbar^2}\bigg[\frac{\Phi_{BES}(T)}{B_hT} +
\frac{\Phi_{FDS}(T,\lim_{\Delta_{SG}\to 0})}{B_sT^{4/3}} \nonumber
\\&& ~~~~+ \gamma\frac{\Phi_{iFDS}(T)}{A_eT^2}\bigg]. \label{eq:5}
\end{eqnarray}

Notice that the resistivity term in Eq.~(\ref{eq:5}) originates
from the FCSS liquid, $\rho_{ab} = \sigma^{-1}_s + \sigma^{-1}_h +
\sigma^{-1}_{s + h \to e} = \sigma^{-1}_s + \sigma^{-1}_h +
\gamma\sigma^{-1}_e = \rho_{ab} + \gamma\rho_c$ that have been
derived in Ref.~\cite{arulsamy3,arulsamy5}. the term
$\sigma^{-1}_{s + h \to e}$ is defined to be the resistivity
caused by the process $s$ + $h$ $\to$ $e$ occurring in the
$ab$-planes. Alternatively, it is the resistivity caused by the
electrons in $ab$-planes. If $s$ + $h$ $\to$ $e$ is completely
blocked in $ab$-planes then $\sigma^{-1}_{s + h \to e}$ = 0. Any
increment in $\sigma^{-1}_e$ also increases $\sigma^{-1}_{s + h
\to e}$ therefore $\sigma^{-1}_{s + h \to e}$ =
$\gamma\sigma^{-1}_e$. In addition, the stated resistivity also
neglects the spinon-pairing contribution, as required by the heat
conductivity equation derived earlier in the absence of
spinon-pairing effect. $B_{LH}$ and $A_{LH}$ are holon and
electron's scattering rate dependent constants (independent of
$T$) in the vicinity of both {\bf H} and $\nabla T$. The
additional $T$ contribution is noted with $<T>$
and~\cite{arulsamyi} $n_e \approx \sqrt{np}=
\frac{k_BT}{\pi\hbar^2}(m_e^*m_h^*)^{1/2}\exp(\frac{-E_I}{k_BT})$
in which, $n_h$ and $n_e$ are the holon and electron's
concentration respectively. Note that the ionization energy
($E_I$) denotes the electron's charge Pseudogap ($\Delta_{PG}$).
In accordance with Refs.~\cite{arulsamy5,anderson6}, there are two
types of scattering rates with respect to $T$-dependence namely,
$\tau_{tr}$ $\propto$ $1/T$ and $\tau_{H}$ $\propto$
$\tau_{tr}/T$. Large phase space of $1/T$ is required for
$\tau_{H}$ due to spinon-holon pair scattering in the presence of
{\bf H} since both spinon and holon are at the mercy of {\bf
H}$\times${\bf E} unlike in the presence of electric field ({\bf
E}) only. Now, considering the pure CSS liquid, one can show that
Eq.~(\ref{eq:5}) can be reduced to

\begin{eqnarray} &L^{(ab)}_H& =
\frac{m_hB_{LH}E_F}{e^2n_h\pi\hbar^2}\bigg[\frac{\Phi_{BES}(T)}{B_h}
+ \frac{\Phi_{FDS}(T)}{B_sT^{1/3}}\bigg]. \label{eq:6}
\end{eqnarray}

At $T_{crossover}$ $>$ $T_c$, the linearity of $L^{(ab)}_H(T)$ for
pure CSS liquid will be inadequate as a result of $T_{crossover}$
($\Delta_{PG}$) effect from $\rho_{ab}(T)$. Recall that both
$\kappa_{ab}(T)$ and $L^{(ab)}_H(T)$ are unsuitable below $T^*$ as
explained earlier. Anyway, all the derivations thus far contain no
uncontrolled approximations or any additional assumptions, apart
from the ones stated here and in Ref.~\cite{arulsamy5}.

\section{3. Analysis}

The calculated heat capacity in $ab$-planes, $\mathcal{C}_{ab}(T)$
is found to decrease linearly with $T$ above $T_c$ (normal state).
In the case of $E_F$ $>$ $T_c$, there is a crossover in
$\mathcal{C}_{ab}(T)$ and note that it does not represent the heat
capacity jump due to Cooper pairs and/or condensed Bosons because
this crossover is an exponential one without a drastic jump.
Figure~\ref{fig1}a)-d) depict the variation of $\mathcal{C}_{ab}$
with $T$ for both FCSS and CSS liquid. It is worth noting that
both $T_{crossover}$ and $T^*$ tend to deviate
$\mathcal{C}_{ab}(T)$ downward from $T$-linear opposing the
exponential increase due to $E_F$. This is because of the
proportionality, $\mathcal{C}_{ab}(T)$ $\propto$ $\exp[E_F/k_BT] +
\exp[(E_F-E_I)/k_BT] + \exp[(E_F-\Delta_{SG})/k_BT]$.
Interestingly, the indirectly measured electronic heat capacity by
Loram {\it et al}.~\cite{loram} at $T$ $>$ $T_c$ (with $T$-linear
property and a slight upward curve near $T_c$, see Fig. 10 in
Ref.~\cite{loram}) is identical with the calculated
$\mathcal{C}_{ab}(T)$. The parameters used to obtain those plots
are also given in the figure itself. Of course, $E_I$ is absent in
Fig.~\ref{fig1}c)-d) since the latter two plots are for pure CSS
liquid. Interestingly, those results were obtained regardless of
the statistics used, FDS or BES since both Fermions and Bosons
carry entropy identically at higher temperatures, above $E_F$.
Simply put, large $E_F$ suppresses the ability of both Fermions
and Bosons to transfer heat and instead gives rise to the
exponentially-increasing heat capacity below $E_F$. Notice that
all energies were given in Kelvin (K).

The calculated plots for the heat conductivity, $\kappa_{ab,c}(T)$
are given in Fig.~\ref{fig2}a)-d). Those curves remained inversely
proportional to $T$, regardless of the reasonable magnitudes of
$E_F$ (for CSS curve only) or both, $E_F$ and $E_I$ (for FCSS
curve only). Mathematically, the parameter, $E_F$ will give rise
to exponential increase of $\kappa_{ab,c}(T)$ if $E_F$ $>$ $T_c$
whereas, $E_I$ leads to an opposite effect in which $E_I$ will
slightly reduce the rate of which $\kappa_{ab,c}(T)$ increases
with reduction in $T$ if $E_I$ $>$ $T_c$. In other words, the
above-stated $E_I$-effect will only come into play for certain
underdoped YBa$_2$Cu$_3$O$_{7-\delta}$ compounds that strictly
satisfy the condition $\Delta_{PG}$ ($T_{crossover}$) $>$ $T_c$
(observable from the resistivity versus temperature measurements).
Unlike resistivity, there will not be any observable crossover for
$\kappa_{ab,c}(T)$ since only the magnitude of inverse
proportionality with $T$ will be reduced, without changing the
overall $T$-dependence. As a consequence, $\kappa_{ab,c}(T)$ will
remain inversely proportional to $T$ and the magnitude of this
inverse proportionality with $T$ in $\kappa_{ab,c}(T)$ will be
reduced (reduced heat conductivity) in certain underdoped
YBa$_2$Cu$_3$O$_{7-\delta}$ compounds. Similar to
$\mathcal{C}_{ab}(T)$, $\kappa_{ab,c}(T)$ $\propto$
$\exp[E_F/k_BT] + \exp[(E_F-E_I)/k_BT]$. Recall that spinon
pairings have been omitted ($T^*$ = $\Delta_{SG}$ = 0) for
$\kappa_{ab,c}(T)$.

The Hall-Lorenz ratio has been the most difficult parameter to be
explained by any theory without violating the $T$-dependence of
resistivity, Hall resistance and Hall angle or at least any one of
them. Remarkably, the Hall-Lorenz ratio based on FCSS liquid
(Eqs.~(\ref{eq:5}) and~(\ref{eq:6})) after incorporating the
$\mathcal{C}_{ab}(T)$, $\kappa_{ab,c}(T)$ and $\rho_{ab}(T)$
reproduces the experimental $T$-linear property for overdoped
YBa$_2$Cu$_3$O$_{7-\delta}$, as long as $T_{crossover}$ ($E_I$
from the resistivity) $<$ $T_c$. Note that the $E_I$ from the heat
conductivity will deviate $L_H^{(ab)}$ downward while the $E_I$
that originates from the resistivity will give an upward
exponential deviation. Such behavior can be trivially verified
from these equations, Eqs.~(\ref{eq:2}),~(\ref{eq:4})
and~(\ref{eq:5}). Actually, $T_{crossover}$ is a characteristic of
certain underdoped YBa$_2$Cu$_3$O$_{7-\delta}$ compounds. The
calculated $T$-linear effect is in excellent agreement with the
phonon-independent experimental Hall-Lorenz ratio
results~\cite{xu}.

It has been repeatedly mentioned that phonons have been neglected
in this work in which, only the electronic version is emphasized
rather than the overall thermal effects. The reason being, to
investigate the $T$-dependence of an entirely electronic
Hall-Lorenz ratio. The Hall-Lorenz ratio that has incorporated the
heat conductivity with phonon contribution also provide the
$T$-linear effect semi-theoretically~\cite{arulsamy5}. It is well
established that the resistivity measurements thus far do not
require phonon inclusion, which can be understood by realizing
that the electron-phonon ($e$-$ph$) coupling observed via ARPES
technique by Lanzara {\it et al}.~\cite{lanzara} actually supports
the notion of polaronic effect above $T_c$ in cuprates. One should
note that $e$-$ph$ coupling does not mean that there is a $e$-$ph$
scattering since normal state $\rho(T)$ measurements thus far
failed to reveal any $e$-$ph$ scattering (strong $T$-dependence).
Actually, this is not because of $\rho(T)$'s blindness, but due to
polarons represented by ionization energy, which gives rise to
effective mass of electrons instead of strong $T$-dependence. The
heavier $m^*$ implies the existence of polarons in the normal
state of HTSC that also suppresses $e$-$ph$ scattering but not
$e$-$ph$ coupling in term of polaronic effect. Similarly, isotope
effect ($^{18}$O, $^{16}$O) in cuprates~\cite{hofer,iyo,keller}
also reinforces the polaronic contribution via $e$-$ph$ coupling
rather than $e$-$ph$ scattering. Furthermore, Bloch-Gr\"{u}neisen
formula has been utilized lately~\cite{arulsamy5} to justify
convincingly that $e$-$ph$ scattering is not applicable for the
resistivity and Hall-Lorenz ratio of YBa$_2$Cu$_3$O$_{7-\delta}$.

\section{4. Conclusions}

In conclusion, the temperature dependencies of
thermo-magneto-electronic properties such as heat capacity, heat
conductivity and Hall-Lorenz ratio, for
YBa$_2$Cu$_3$O$_{7-\delta}$ has been derived using both electron
contaminated CSS and pure CSS liquid. The theoretical plots
discussed based on this liquid also shown to be applicable in the
experimental thermo-magneto-electronic transport properties, apart
from the magneto-electronic transport properties, which have been
published previously. Phonon contribution has been neglected
throughout the derivations while spinon-pairing is dropped for the
derivation of heat conductivity and Hall-Lorenz ratio parameters.
The latter is unavoidable since the influence of $\nabla T$ on
paired spinons and eventually on $\tau_{ab}$ is unclear. The
assumptions throughout these derivations are again consistently
based on Anderson's hypothesis ($\tau_{tr}$ $\propto$ $1/T$ and
$\tau_{H}$ $\propto$ $\tau_{tr}/T$), the Ioffe-Larkin rule
($\sigma^{-1}$ = $\sigma_s^{-1}$ + $\sigma_h^{-1}$) and FCSS
liquid ($\sigma^{-1}$ = $\sigma_s^{-1}$ + $\sigma_h^{-1}$ +
$\sigma_e^{-1}$).

\section*{Acknowledgments}
ADA is grateful to Arulsamy Innasimuthu, Sebastiammal Innasimuthu,
Arokia Das Anthony and Cecily Arokiam, the financiers of the
Condensed Matter Group-Ampang, for setting up the computing
facilities.

\begin{figure}
\caption {Theoretical plots for the heat capacity in $ab$-planes,
$\mathcal{C}_{ab}(T)$ above 90 K are calculated using
Eq.~(\ref{eq:1}). The calculated plots are obtained at different
$E_I$, $T^*$ and $E_F$. The plots in a) and b) are for FCSS liquid
while c) and d) are for pure CSS liquid ($m_e^*\Phi_{iFDS}(T)$
$\to$ 0). In a) and b), $\mathcal{C}_{ab}$ with $T$ linear
property dominates as long as $E_F$ $<$ $T_c$ in which, an
exponential increase takes over below $E_F$ if $E_F$ $>$ $T_c$
and/or $E_F$ $>$ $E_I,T^*$. This latter scenario is due to the
fact that both $T^*$ and $E_I$ tend to recover the $T$-linear
effect, opposing $E_F$. The plots in c) and d) indicate the
similarities between CSS and FCCS liquid since $E_F$ is the only
parameter in $\mathcal{C}_{ab}(T)$ that is responsible for the
large exponential deviation from the $T$-linear effect. Actually,
both $T^*$ and $E_I$ will give rise to a slight downward deviation
from the $T$-linear property.} \label{fig1}
\end{figure}

\begin{figure}
\caption {The phonon independent heat conductivity in $ab$-planes,
$\kappa_{ab}(T)$ of both FCSS and CSS liquid have been plotted in
a), b) and c), d) respectively. As anticipated, the variation of
$\kappa_{ab}(T)$ with respect to $T$ is weakly inversely
proportional to $T$ for all reasonable $E_F$s and $E_I$s.
Consequently, $\kappa_{ab}(T)$ from both FCSS and CSS liquid
shares a similar trend against $T$. The equations used to obtain
the FCSS and CSS plots are Eqs.~(\ref{eq:2}) and~(\ref{eq:3})
respectively.} \label{fig2}
\end{figure}

\begin{figure}
\caption {The Hall-Lorenz ratio in $ab$-planes, $L^{ab}_H$ gives a
convincing $T$-linear effect solely as a result of charge and spin
separation. Exponential deviation from this linearity could arise
from either $E_I$ and/or $E_F$ depending whether $E_I$ $>$ $E_F$
$>$ $T_c$ or $T_c$ $<$ $E_I$ $<$ $E_F$. The theoretical plots for
FCSS and CSS liquid, neglecting spinon-pairing are given in a), b)
and c), d) respectively. Equation~(\ref{eq:5}) for FCSS and
Eq.~(\ref{eq:6}) for pure CSS liquid have been employed.}
\label{fig3}
\end{figure}

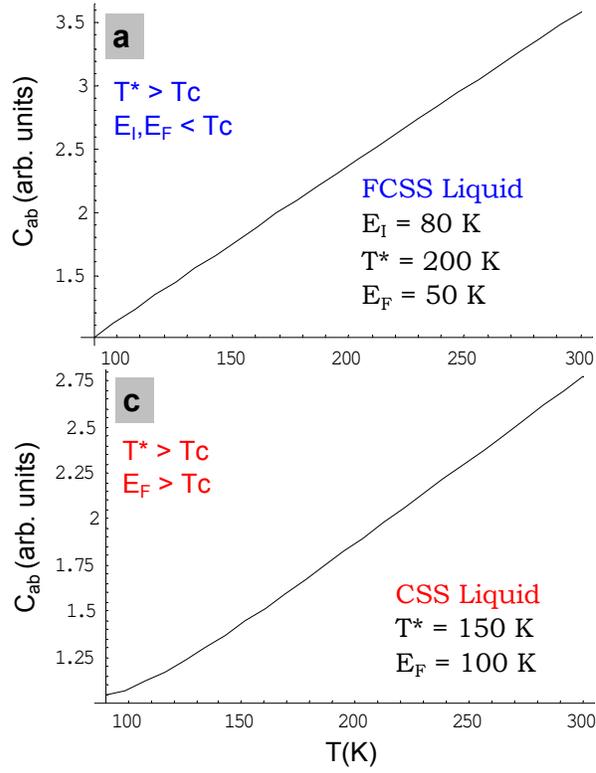
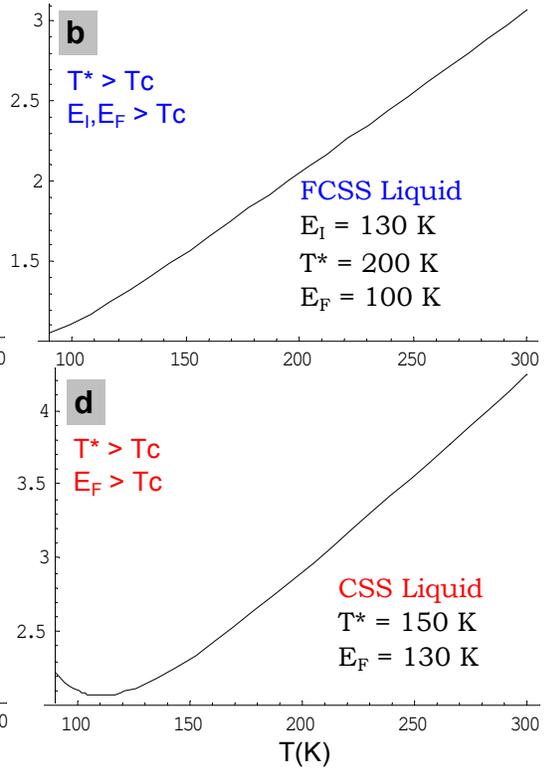
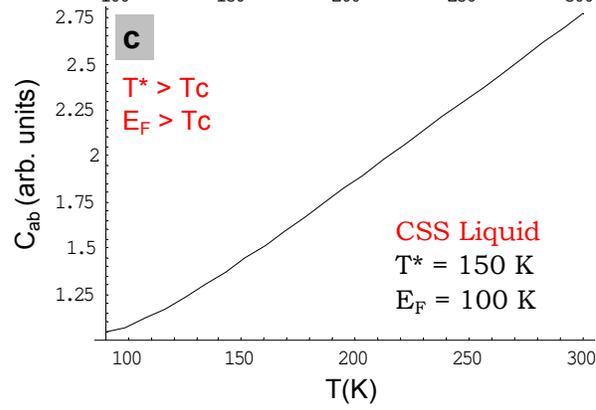
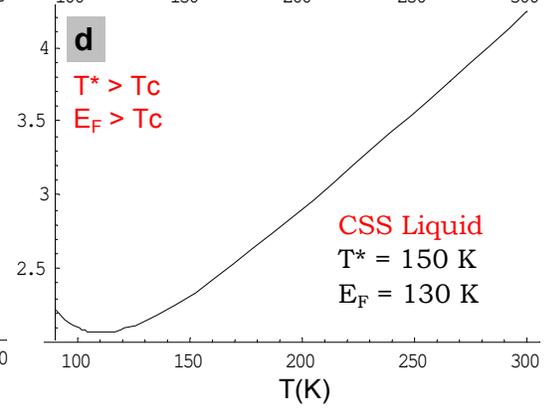
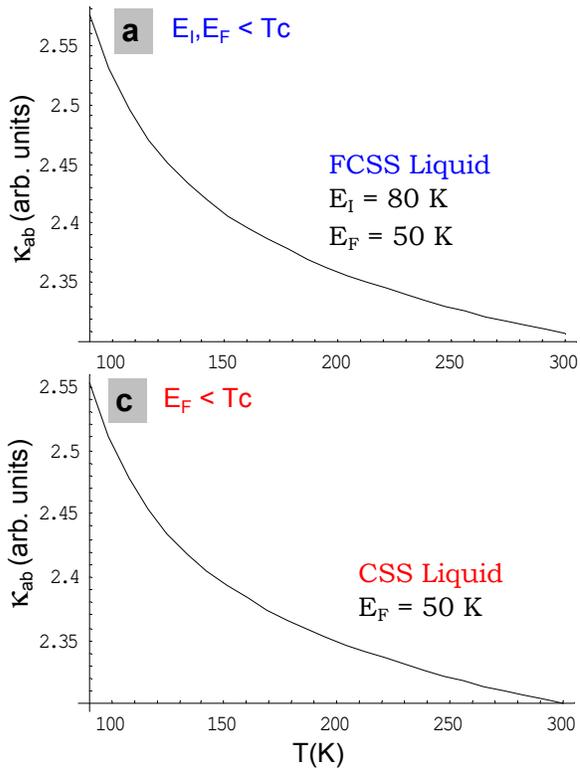
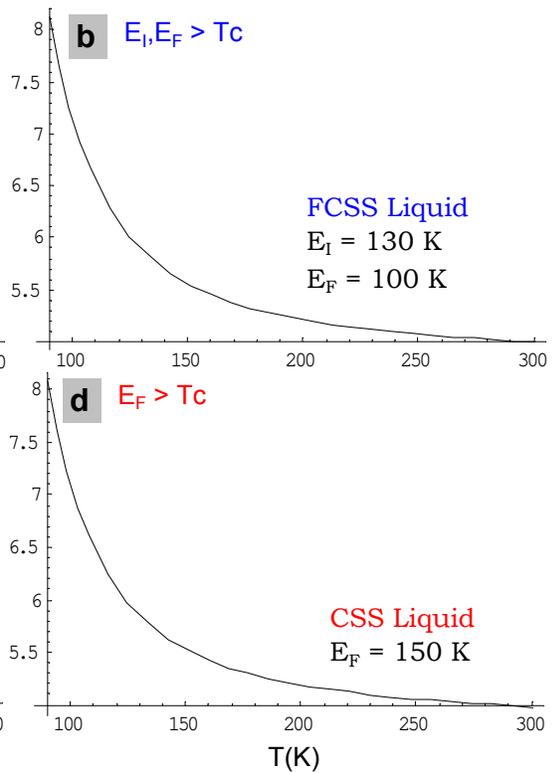
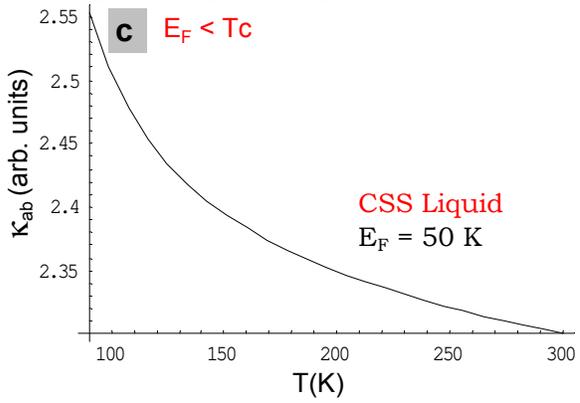
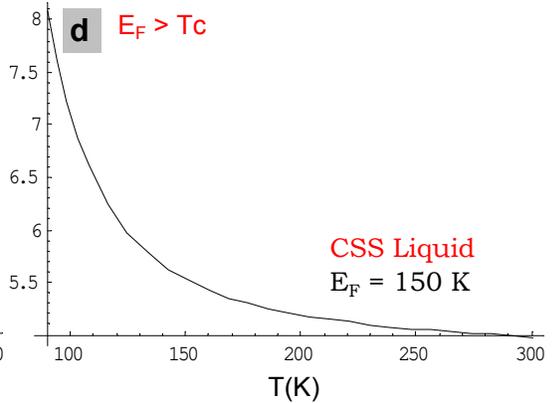
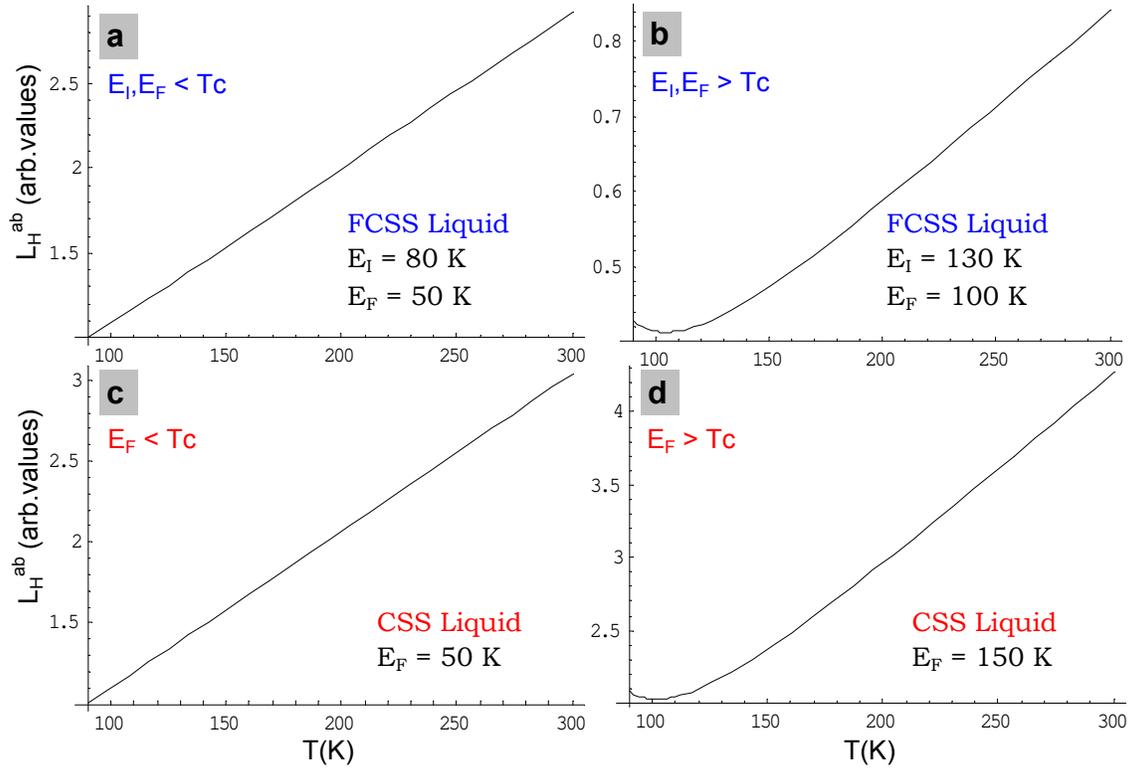
\end{document}